\documentclass[12pt,a4paper]{article}
\usepackage{graphicx}
\begin{document}
 
\thispagestyle{empty}
 
\title{Group actions, geodesic loops, and symmetries of compact hyperbolic 
3-manifolds.}
\author{Peter Kramer,\\
Institut f\"ur Theoretische Physik der 
Universit\"at  72076 T\"ubingen, Germany}
\maketitle

\section*{Abstract.}
Compact hyperbolic 3-manifolds are used in cosmological models. 
Their topology is characterized by their homotopy group $\pi_1(M)$
whose elements multiply by path concatenation.
The universal covering of the compact manifold $M$ 
is the hyperbolic space
$H^3$ or the hyperbolic ball $B^3$. 
They share with $M$ a Riemannian metric 
of constant negative curvature and allow for the isometric action
of the group $Sl(2,C)$.
The homotopy group $\pi_1(M)$
acts as a uniform lattice $\Gamma(M)$ on $B^3$ and tesselates it by
copies of $M$.
Its elements $g$ produce preimage and image points for geodesic sections
on $B^3$ which by self-intersection form geodesic loops  on $M$. 
For any fixed hyperbolic  $g \in \Gamma$   
we construct a continuous commutative two-parameter normalizer  
$N_g <Sl(2,C)$ and its
orbit surfaces on $B^3$. The orbit surfaces classify  sets of
geodesic loops of equal length. We give general
expressions for the length of geodesic loops and for
the defect angle at the self-intersection on $M$ in terms of 
the group parameters of $g$ and orbit parameters on $B^3$.
Geodesic loops of minimal
length, given  from the
character $\chi(g)$, belong to a single orbit. These and only 
these minimal
geodesic loops have vanishing defect angle and hence are smooth
everywhere.
The role of symmetries is illuminated by the example
of the dodecahedral hyperbolic Weber-Seifert manifold $M$.
$\Gamma(M)$  is normal in the hyperbolic Coxeter group
with Coxeter  diagram 
${\bf \circ \frac{5}{}\circ\frac{3}{}\circ\frac{5}{}\circ\;}$.
This leads to symmetry relations between geodesic loops.

\section{Introduction.}

Models of a closed cosmos with nontrivial topology were reviewed by 
Lachieze and Luminet in \cite{LA} and by Levin in \cite{LE}. 
The predictions  from the models may  
be compared directly with  astronomical data, compare Fagundes \cite{FAG1}, 
\cite{FAG2},
or with the  autocorrelation 
of the observed cosmic mass density \cite{LA} pp. 200-201.
A well-known example is the dodecahedral hyperbolic manifold $M$  due 
to  Weber and Seifert \cite{WE},
compare Best \cite{BE} and Thurston \cite{TH} pp 36-37.
Closed hyperbolic manifolds can be ordered by their volume.
Those of small volume due to Thurston and to Weeks
have found particular attention in cosmology \cite{LE} p. 265.
As pointed out in \cite{LE} p. 266, geodesic loops 
on the manifold $M$ convey important information for the spacing
of ghost images and for the autocorrelation of the mass density.

We use continuous and discrete groups for the analysis of geodesic loops.
We consider a general compact hyperbolic manifold $M$ whose universal covering 
is the hyperbolic space $H^3$ or hyperbolic ball $B^3 \sim H^3$.
$H^3$ is equivalent
to the coset space $SO_{\uparrow}^{+}(1,3,R)/SO(3,R)$, admits the isometric action of 
$SO_{\uparrow}^{+}(1,3,R)$ or of its universal covering group 
$Sl(2,C)$, and has constant negative curvature \cite{WO}.
The Minkowski metric restricted to $H^3$ yields the notions of
geodesics and their length, which carry over to $M$.
The homotopy group $\pi_1(M)$ multiplies by path concatenation. 
Typically  it is a finitely generated (and even finitely presented) 
infinite discrete group. 

The uniform lattice $\Gamma(M)$ is isomorphic to $\pi_1(M)$, acts
as a discrete subgroup of $Sl(2,C)$ without fixpoints
on the universal covering manifold $B^3$,
and generates a tesselation by copies of $M$. 
Pairs of points on different
copies of $M$ in $B^3$ are equivalent if there is an element $g$ of  
$\Gamma(M)$ which has this pair as preimage and image. 

Equivalent
points of $B^3$ are identified on $M$. Any geodesic section on $B^3$
between equivalent points when mapped to $M$ forms a geodesic loop with 
self-intersection.

A classification of geodesic loops on $M$ should deal with  
two aspects: (i) Find the variety of geodesic loops for a 
given fixed element $g \in \Gamma$ . (ii) Compare 
geodesic loops for different elements of $\Gamma(M)$,
leading to a length spectrum of geodesic loops. This requires 
an analysis of the elements of $\Gamma(M)$, taken as words
in its generators.

In what follows we consider mainly the aspect (i) for a general closed

hyperbolic manifold $M$. 
In section 2 we briefly describe the groups, in section 3 the geometry of
the hyperbolic space $H^3$ and hyperbolic ball $B^3$, and the group 
actions. In section 4
we develop the classification of geodesic loops, and in section 5 and 6
we illustrate symmetries on the compact hyperbolic Weber-Seifert manifold.

\section{The Lorentz group $SO_{\uparrow}^{+}(1,3,R)$, its covering, 
and Weyl reflections.}

The universal covering manifold for a compact hyperbolic manifold $M$ 
is the hyperbolic space $H^3$, a homogeneous space under the Lorentz group
$SO_{\uparrow}^{+}(1,3,R)$.
In this section we collect results and notations for the action of this 
and related continuous groups. In the next sections  they will be used 
for the discrete uniform lattice $\Gamma(M)$.

\subsection{$Sl(2,C)$: Class and in-class structure, adjoint representation.}
The universal covering of the proper time-preserving Lorentz
group  $SO_{\uparrow}^{+}(1,3,R)$ is the unimodular group $Sl(2,C)$.
Both groups have $6$ real parameters. 

We omit from the discussion unipotent elements
of $Sl(2,C)$ with a Jordan decomposition since such elements cannot
appear in $\Gamma$ by \cite{ON}, theorem 4.1 (Kazhdan/Margulis), p. 79.
Then all elements $g$ of interest have class representative $g_0$ 
in diagonal form. Exponential parameters
can be used to display $g_0$  as  
\begin{equation}
\label{g1}
g_0=
\left[
\begin{array}{ll}
\exp(c+i\gamma) & 0\\
0& \exp(-(c+i\gamma))
\end{array}
\right],\;
-\infty <  c < \infty,\; 0\leq \gamma < 2\pi.
\end{equation}
with two real class parameters $c,\; \gamma$. 
The elements 
eq. \ref{g1} belong to a subgroup $H$ isomorphic to $SO(1,1,R) \times SO(2,R)$
but in diagonal, not in standard form. We call hyperbolic the 
elements eq. \ref{g1} and their conjugates which in addition obey $|c|>0$.
Given a general element $g \in Sl(2,C)$, we can determine its 
class parameters by use of the complex character or trace $\chi(g)$,
\begin{equation}
\label{g2}
\frac{1}{2} \chi(g_0)= \frac{1}{2}Tr(g_0)= \cosh(c+i\gamma)
= \cosh(c)\cos(\gamma)+i\sinh(c)\sin(\gamma).
\end{equation}
We define the additional elements
\begin{eqnarray}
\label{g3}
g_1&=&g_1(a,\alpha)
=\left[
\begin{array}{ll}
\cosh(a+i\alpha)&\sinh(a+i\alpha)\\
\sinh(a+i\alpha)&\cosh(a+i\alpha)
\end{array}
\right],\; (a, \alpha)\; {\rm real},
\\ \nonumber
g_2&=&g_0(b,\beta)
=\left[
\begin{array}{ll}
\exp(b+i\beta) & 0\\
0& \exp(-(b+i\beta))
\end{array}
\right],\; (b,\beta)\; {\rm real}.
\end{eqnarray}
By use of complex Euler angles it can be shown that the products $g_2g_1$ parametrize the 
cosets $Sl(2,C)/H$.
By $\sigma_1,\sigma_2,\sigma_3$ we denote the standard hermitian 
Pauli matrices and by $e$ the $2 \times 2$ unit matrix.
We claim: The general element $g$ of the class with representative eq. \ref{g1}
may be written as
\begin{eqnarray}
\label{g4}
g&=& (g_2g_1) g_0 (g_1g_2)^{-1}, 
\\ \nonumber 
&=&
\cosh(c+i\gamma) + \sinh(c+i\gamma)
\sum_{i=1}^3 \eta_i \sigma_i,\; \sum_{i=1}^3(\eta_i)^2=1,
\\ \nonumber 
\left[
\begin{array}{l}
\eta_1\\
\eta_2\\
\eta_3
\end{array}
\right]
&=& Ad(g_2g_1)
\left[
\begin{array}{l}
0\\
0\\
1
\end{array}
\right],
\\ \nonumber
Ad(g_2g_1)
 &=&\left[
\begin{array}{lll}
\cosh(2(b+i\beta))&-i\sinh(2(b+i\beta))&0\\
i\sinh(2(b+i\beta))&\cosh(2(b+i\beta))&0\\
0&0&1 
\end{array}
\right]
\\ \nonumber
 &\times&\left[
\begin{array}{lll}
1&0&0\\ 
0&\cosh(2(a+i\alpha))&-i\sinh(2(a+i\alpha))\\
0&i\sinh(2(a+i\alpha))&\cosh(2(a+i\alpha))\\
\end{array}
\right].
\end{eqnarray}
The trace is not affected by the conjugation in eq. \ref{g4} and so
the class parameters $c,\gamma$ are obtained from the character
$\chi(g)=\chi(g_0)$ as in eq. \ref{g2}.
We call $a,\alpha,b,\beta$ the $4$ in-class parameters of the class.
For $a=b=c=0$, eq. \ref{g4}  becomes a variant of the familiar
parametrization of $SU(2)$.

The adjoint representation $Ad(Sl(2,C))$  
is formed by the complex orthogonal
group $SO(3,C)$. For the elements $g_2,g_1 \in Sl(2,C)$, the adjoint 
representation is generated by the matrices $Ad(g_2),\; Ad(g_1)$ in eq. \ref{g3}.
It is  expressed by two complex rotations which
extend the  two real rotations used with the adjoint representation 
of $SU(2)$. For $a=b=c=0$, the adjoint representation becomes equivalent  to
$Ad(SU(2))\sim SO(3,R)$. The homomorphism $Sl(2,C) \rightarrow Ad(Sl(2,C))$
is two-to-one with $Ad(g)=Ad(-g)$.

\subsection{The proper time-preserving Lorentz group and Weyl reflections.}

We obtain in the usual fashion the two-to-one homomorphism from $Sl(2,C)$ 
to the Lorentz group
with transformations
$L(g) \in S0_{\uparrow}^{+}(1,3,R)$ . We
use in Minkowski space $M(1,3)$ the coordinates 
$(x_0,x_1,x_2,x_3)$. The
scalar product is taken as
\begin{equation}
\label{g5}
\langle x,y\rangle = x_0y_0-x_1y_1-x_2y_2-x_3y_3.
\end{equation}
We introduce 
the $2 \times 2$ hermitian matrix
\begin{equation}
\label{g6}
\tilde{x} = x_0e+\sum_{i=1}^3 x_i\sigma_i.
\end{equation}
with $det(\tilde{x})=\langle x,x\rangle$.
The linear Lorentz action $L(g)=L(-g)$  of $Sl(2,C)$ on 
$M(1,3)$ is
\begin{eqnarray}
\label{g7}
g \in Sl(2,C): \tilde{x} \rightarrow \tilde{x}'&=& g\tilde{x}g^{\dagger}, 
\\ \nonumber
x_{\mu}' &=& \sum_0^3 L_{\mu\nu}(g) x_{\nu}.
\end{eqnarray}
In particular one finds from eqs. \ref{g1}, \ref{g3}, \ref{g7}
\begin{eqnarray}
\label{g8}
L(g_0(c,\gamma))
&:=& 
\left[
\begin{array}{llll}
\cosh(2c) &0&0& \sinh(2c)\\
0&\cos(2\gamma) & \sin(2\gamma)&0\\
0& -\sin(2\gamma) & \cos(2\gamma)&0\\
\sinh(2c) &0&0& \cosh(2c)
\end{array}
\right],
\\  \nonumber
L(g_1(a,\alpha))
&:=& 
\left[
\begin{array}{llll}
\cosh(2a) & \sinh(2a)&0&0\\
\sinh(2a) & \cosh(2a)&0&0\\
0&0& \cos(2\alpha) & \sin(2\alpha)\\
0&0& -\sin(2\alpha) & \cos(2\alpha)
\end{array}
\right],
\\  \nonumber
L(g_2)&:=&L(g_0(b,\beta)).
\end{eqnarray}
The Lorentz transformations $L(g_0(c,\gamma))$ for hyperbolic
$g_0$ with $|c|>0$ as defined after eq. \ref{g1} can have no fixpoint on the 
coset space $H^3$. 

Given a vector $k \in M(1,3),\; \langle k,k\rangle\neq 0$, 
we define the Weyl reflection operator 
$W_k: M(1,3) \rightarrow M(1,3)$ by 
\begin{equation}
\label{g9}
W_k: x \rightarrow x-2\frac{\langle k,x\rangle}{\langle k,k\rangle}\; k.
\end{equation}
The Weyl reflection preserves any point of the reflection hyperplane
$\langle x:\; \langle k,x\rangle=0 \rangle$. 
Weyl operators are isometries with respect to the metric of $M(1,3)$.
They allow to extend the proper time-preserving Lorentz group 
by space reflections. In particular for $k=e_2=(0,0,1,0), \langle k,k\rangle=-1$, the Weyl reflection
$W_{e_2}$ inverts only the coordinate $x_2$. Under Lorentz transformations
one easily proves the conjugation law
\begin{equation}
\label{g10}
L(g)W_kL(g^{-1})=W_{L(g)k}.
\end{equation}

\section{The hyperbolic space $H^3$ and ball $B^3$.}

The hyperbolic space $H^3$ arises as the universal covering space of 
compact hyperbolic manifolds.  This universal covering space plays 
a key role in the analysis of geodesic loops in sections 4-6.
For details on the hyperbolic space and ball we refer to 
Ratcliffe \cite{RA} pp. 56-104 and pp. 127-135 respectively.

The hyperbolic space is the coset space 
$H^3:= SO_{\uparrow}^{+}(1,3,R)/SO(3,R)$.
In $M(1,3)$ its points form the hyperboloid 
\begin{equation}
\label{g11}
H^3= \langle x|\langle x,x \rangle =1, x_0 \geq 1\rangle.
\end{equation}
We follow \cite{RA} up to a sign and rewrite the scalar product 
eq. \ref{g5} in $M(1,3)$
as
\begin{eqnarray}
\label{g12}
\langle x,y\rangle &=& x_0y_0-(x,y),\\
\nonumber
(x,y)&:=& x_1y_1+x_2y_2+x_3y_3.
\end{eqnarray} 
Given two points $x,y \in H^3$, we take their scalar product as
the restriction of the scalar product
$\langle x,y \rangle$ from $M(1,3)$ to $H^3$.
The restricted scalar product on $H^3$ is positive definite.
$H^3$ with this metric can be shown to be a space of 
constant negative curvature \cite{WO} .

The model of the conformal ball $B^3$ for $H^3$ is obtained from 
the points of the hyperboloid eq. \ref{g11}
by the fractional linear map to a Euclidean space $E^3$,
\begin{eqnarray}
\label{g13}
x_i \rightarrow \xi_i&=&  \frac{x_i}{1+x_0},\; i=1,2,3,\; (\xi,\xi) <1,
\\ \nonumber 
x_i&=& \frac{2\xi_i}{1-(\xi,\xi)},\; i=1,2,3.
\end{eqnarray}
Although all points of $H^3$ from eq. \ref{g13} map bijectively 
into points of $B^3$, we shall need points in $E^3$ but outside $B^3$
to characterize its symmetries.
We shall use the scalar product $(,)$ with respect to the 
space-like components of vectors from $M(1,3)$, for vectors 
in $B^3$, and in $E^3$  
embedding $B^3$. Consider a space-like vector 
$k \in M(1,3),\; k_0\neq 0,$ 
and a point $x: \langle x,x\rangle =1$ of $H^3$.
Using for $x$ the coordinates eq. \ref{g13} on $B^3$ one finds
\begin{eqnarray}
\label{g14}
\langle k,x\rangle &=&k_0(1+x_0)\frac{1}{2}(1+(\xi,\xi)-2(q,\xi)),
\\ \nonumber
k &\rightarrow& q=k_0^{-1}(k_1,k_2,k_3).
\end{eqnarray}
{\bf 1 Lemma}: The intersection of the hyperplane 
$\langle k,x\rangle=0,\; k_0 \neq 0$ in
$M(1,3)$ with the hyperboloid $H^3$ in the conformal ball model 
becomes the intersection of a M\"obius 
sphere $S^2(q,R)$  of center $q, (q,q)>1$ and radius $R$ with $B^3 \subset E^3$, 
\begin{eqnarray}
\label{g15}
(\xi-q,\xi-q)-R^2&=&0,
\\ \nonumber
q&=& k_0^{-1}(k_1,k_2,k_3),\; R=\sqrt{(q,q)-1}.
\end{eqnarray}
The M\"obius sphere has orthogonal  intersections with $\partial B^3$.

{\em Proof}: It suffices to rewrite part of eq. \ref{g14} as
\begin{equation}
\label{g16}
(1+(\xi,\xi)-2(q,\xi))=(\xi-q,\xi-q)-(q,q)+1=(\xi-q,\xi-q)-R^2.
\end{equation}
The condition $R^2=(q,q)-1$ assures that the M\"obius sphere $S^2(q,R)$ 
has orthogonal intersections with the surface 
$\partial B^3:\; (\xi,\xi)=1$. The sphere  $S^2(q,R)$  separates the points of $B^3$
into two disjoint parts.
In case of a vector $k: (0,k_1,k_2,k_3)$ when $q$ in eq. \ref{g15}
is not well-defined we replace the sphere in $E^3$ by a 
plane through the origin and perpendicular to $(k_1,k_2,k_3)$.

Given  a Weyl reflection eq. \ref{g9} with space-like Weyl vector $k$
one finds

{\bf 2 Lemma}: A Weyl reflection $W_k$ eq. \ref{g9}  with Weyl vector $k$ space-like 
 in $E_3$ becomes a Möbius inversion in the sphere $S^2(q,R)$ with center and radius 
$(q,R)$ from eq. \ref{g15}. This Möbius inversion sends $B^3$ into $B^3$
and transforms points inside and outside of $S^2(q,R)\cap B^3$ into one another.
The Möbius inversion is conformal, angles between geodesics are 
preserved.\\
{\em Proof}: We can write the action of the Weyl operator in $B^3$ 
from eqs. \ref{g9}, \ref{g13} as
\begin{equation}
\label{g17}
\xi_i \rightarrow \kappa_i= 
\frac{\xi_i+(1+(\xi,\xi)-2(q,\xi))k_0^2q_i}{1+(1+(\xi,\xi)-2(q,\xi))k_0^2},
\; i=1,2,3.
\end{equation}
If in $E^3$ we transform to new coordinates with respect to the 
center $q$ of $S^2(q,R)$ by putting in eq. \ref{g13} $\xi_i=q_i+u_i,\;
\kappa_i= q_i+v_i$, we get the Weyl reflection eq. \ref{g9}
on $B^3$ in the form
\begin{equation}
\label{g18}
u_i \rightarrow v_i= 
u_i \frac{R^2}{(u,u)},\; (v,v)(u,u)=R^4.
\end{equation}
This is the standard form of a Möbius inversion in the sphere $S^2(q,R)$,
compare \cite{RA} pp.109-11,   and therefore is conformal.

\section{Geodesics on $H^3$ and $B^3$.}

As pointed out in the introduction, for any geodesic loop on 
a fixed compact hyperbolic manifold $M$
there is a unique element $g \in \Gamma(M)$  acting on $H^3$. 
We wish to characterize the variety of geodesic loops associated with a fixed $g$. To this purpose 
we construct  the continuous normalizer of $g \in SL(2,C)$ and 
its orbits on $H^3$. The orbits classify geodesic loops
by length and direction and yield explicit
expressions for them. A geodesic which closes on $M$ must intersect itself.
From $H^3$ we compute the defect angle at the self-intersection. 
For given $g$, there is a unique set of geodesic loops 
which have minimum length and are smooth, i.e. have vanishing defect angle.

\subsection{Two-parameter normalizer subgroups of $Sl(2,C)$
and orbit surfaces.}

With a discrete element $g$  of $Sl(2,C)$ we associate a continuous
group which allows to classify geodesic sections related to
$g$.   

{\bf 3 Def}:
Consider a discrete  element $g \in Sl(2,C)$ in diagonal
form, $g=g_0(c,\gamma)$, eq. \ref{g1}
with fixed exponential parameters $c,\gamma$.
Define the group $N_{g_0}$ with elements 
\begin{equation}
\label{g19}
h(\lambda,\phi):=g_0(\lambda c/2,\lambda\gamma/2+\phi)
,\; -\infty < \lambda< \infty,\; 0\leq \phi<2\pi.
\end{equation}
This  is the two-parameter 
commutative normalizer $N_{g_0}$ (which coincides with the centralizer) 
of the group generated by $g_0$.
Its elements must commute with $g_0$, eq. \ref{g1}. 

$N_{g_0}$ is isomorphic to
$SO(1,1,R) \times SO(2,R)$. 
With $L(g_0(\lambda c/2, \lambda \gamma/2+\phi))$ we represent
$N_{g_0}$ by Lorentz transformations. The reasons for
our choice of parameters will appear from the 
actions.

Turn to the action of $N_{g_0}$ on $M(1,3), H^3, B^3$.
On $H^3$ we can choose the hyperplane 
$\langle e_3, x\rangle=0$ for the orbit representatives
on $H^3$ under $N_{g_0}$. We incorporate the second parameter 
$\phi$ of $N_{g_0}$ into the representatives and write them 
as
\begin{equation}
\label{g20}
x(\rho,\phi)=(\cosh(2\rho), \sinh(2\rho)\cos(2\phi),
\sinh(2\rho)\sin(2\phi), 0),\;  0\leq \rho <\infty, 0\leq 2\phi<2\pi.
\end{equation}

The points on $H^3, B^3$ under the action of $N_{g_0}$ 
we call orbit surfaces. Each orbit surface is determined by a 
single value of $\rho$.
We call orbit lines the points from the action of 
$N_{g_0}$ 
on $B^3$ for varying
$\lambda$ and fixed $\phi$. The full orbit surface is obtained by 
rotating this orbit line with the angle $\phi$.

{\bf 4 Lemma}: Preimages on a fixed orbit surfaces under $g_0$
are mapped into images on  the same orbit surface. The 
relative geodesic distance between preimage and image point is
independent of the starting point on the orbit surface.  

{\em Proof}: The first part follows from the fact that
$N_{g_0}$ commutes with $g_0$. Any point on an orbit surface may
be written as $L(g_0(\lambda c/2,\lambda\gamma/2))x(\rho,\phi)$. With 
$g_0=g_0(c,\gamma)$ we obtain from 
the scalar product for the geodesic distance between 
preimage and image points under $L(g_0(c,\gamma))$
\begin{eqnarray}
\label{g21}
&&\langle L(g_0(-\lambda c/2,-\lambda\gamma/2))x(\rho,\phi), 
L(g_0(c,\gamma))L(g_0(-\lambda c/2,-\lambda \gamma/2))x(\rho,\phi) \rangle
\\ \nonumber
&& = \langle x(\rho,\phi), 
L(g_0(c,\gamma))x(\rho,\phi) \rangle
\; = \langle x(\rho,0), 
L(g_0(c,\gamma))x(\rho,0) \rangle.
\end{eqnarray}

The Lorentz invariance of the scalar product and
the commutativity of $N_{g_0}$ are crucial for this result.

We choose as specific preimages on $H^3$ for the orbit lines the points 
$L(g_0(-c/2,-\gamma/2))x(\rho,\phi)$.
The images  then are of the form $L(g_0(c/2,\gamma/2))x(\rho,\phi)$.
Preimages and images 
correspond to parameter values $\lambda= \mp 1$ in eq. \ref{g20} respectively.
For fixed $\lambda$, all orbit lines
pass the hyperplane $\langle k,x\rangle =0,
\; k=L(g_0(\lambda c/2, \lambda \gamma/2))e_3$ of $H^3$. 

On $H^3$ the orbit lines take the form
\begin{equation}
\label{g22}
L(g_0(\lambda c/2,\lambda \gamma/2))x(\rho,\phi)
=
\left[
\begin{array}{l}
\cosh(2\rho)\cosh(\lambda c)\\
\sinh(2\rho)\cos(2\phi-\lambda \gamma)\\
\sinh(2\rho)\sin(2\phi-\lambda \gamma)\\
\cosh(2\rho)\sinh(\lambda c)
\end{array}
\right].
\end{equation}

We compute the orbit lines in $B^3$ and find by applying 
eq. \ref{g13} to eq. \ref{g22}
\begin{equation}
\label{g23}
\xi(\lambda)
= (1+\cosh(2\rho)\cosh(\lambda c))^{-1}
\left[ 
\begin{array}{l}
\sinh(2\rho)\cos(2\phi-\lambda \gamma)\\
\sinh(2\rho)\sin(2\phi-\lambda \gamma)\\
\cosh(2\rho)\sinh(\lambda c)
\end{array}
\right].
\end{equation}
For fixed $\lambda$, the orbit lines $\xi(\lambda)$ intersect the 
M\"obius sphere 
$S^2(q,R)$ with 
\begin{equation}
\label{g24}
q(\lambda)=(0,0,{\rm cotanh}(\lambda c)),\; R(\lambda)=(\sinh(\lambda c))^{-1}.
\end{equation}
Since the action of $Sl(2,C)$ is conformal we have

{\bf 5 Lemma}: The orbit lines on $B^3$ for fixed parameter $\lambda$ intersect the
M\"obius sphere eq. \ref{g24}. The angle between tangents to orbit lines 
and  the normal
of the M\"obius sphere depends on the starting value of $\rho$ 
but is independent of $\lambda, \phi$. 

All orbit lines  
for $\lambda=0$ cross the plane $\xi_3=0$.
This choice allows for a simple
visualization of the orbit lines and geodesics in Fig. 1.
The full orbit surfaces are obtained by rotating the orbit lines
to any angle $\phi$ around the $3$-axis.

\begin{center}
\includegraphics{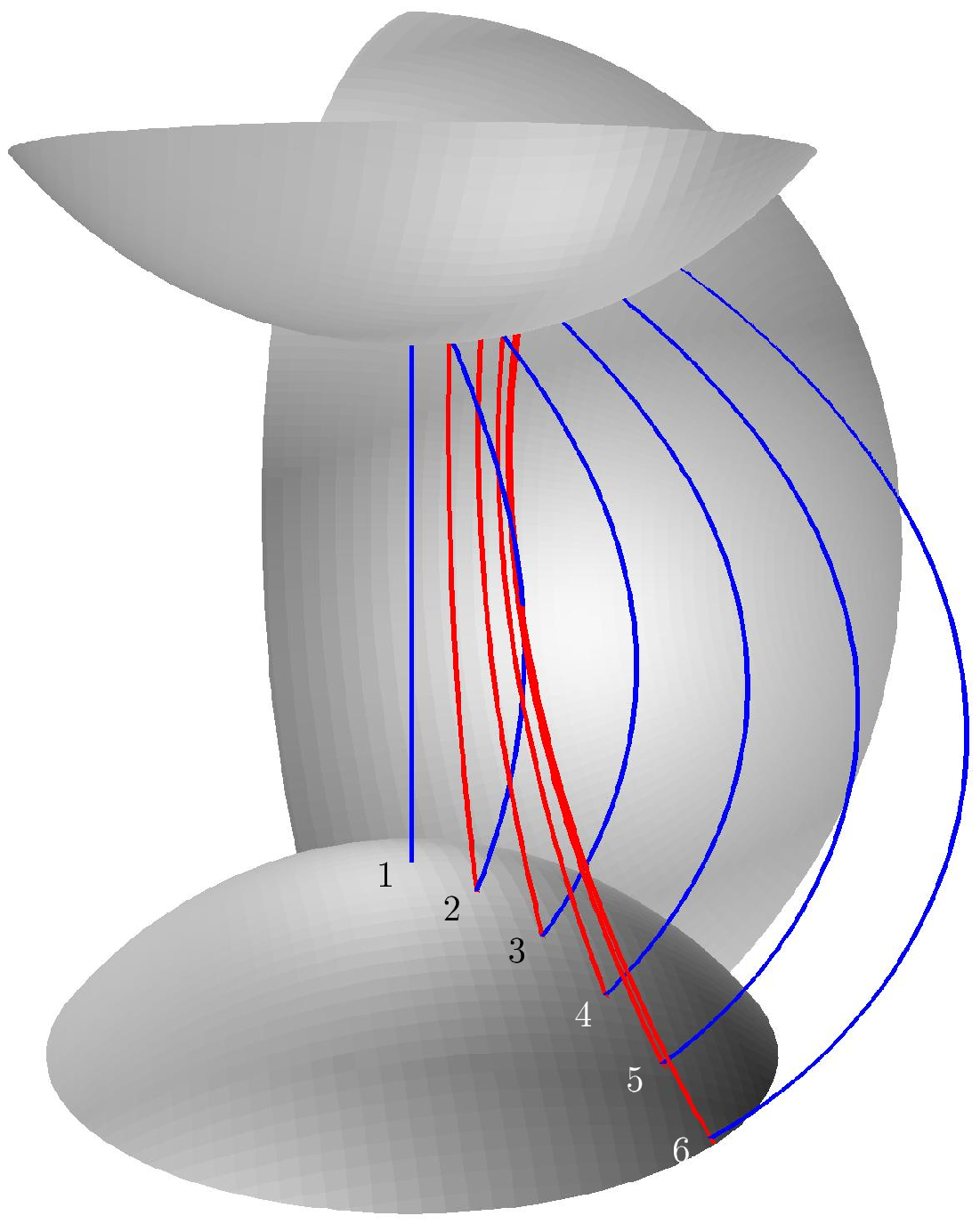}
\end{center}

Fig. 1. A sector of area $\pi$ from the boundary of 
$B^3$ and its intersection with two M\"obius spheres $S^2(q,R)$.
The chosen element $g=C_1 \in \Gamma(M)$ is a generator of $\Gamma$ for the 
compact Weber-Seifert manifold. It maps the lower into the upper
M\"obius sphere. Shown are six orbit lines $1,\ldots ,6$
for this map, and  six geodesic sections  
between preimages
and images under $g= C_1$. The orbit surface arises from each orbit line by a rotation around
the vertical 3-axis.  Only the straight orbit line $1$ coincides 
with the shortest  
geodesic section $1$ under $C_1$. 
\vspace{0.2cm}

\subsection{Geodesic lines on $B^3$.}

{\bf 6 Lemma}: The geodesic connecting two points $\xi, \eta,\; \xi \neq \eta$ on 
$B^3$ is the section between the points
$\langle \xi,\eta \rangle$ on the circle $S^1(q,R)$  which has  two perpendicular intersections with
the surface $\partial B^3$.

We compute the geodesic circle $S^1(q,R)$ explicitly. The conditions that the two points be on the 
same circle with center $q$ and radius $R=\sqrt{(q,q)-1}$ 
in a plane containing $\xi=0$ are
\begin{equation}
\label{g25}
q=\mu_1\xi+\mu_2\eta,\;
2(q,\xi)=(\xi,\xi)+1,\; 2(q,\eta)=(\eta,\eta)+1, .
\end{equation}
Solving these linear equations for $\mu_1,\mu_2$ yields
\begin{equation}
\label{g26}
\left[
\begin{array}{l}
\mu_1\\
\mu_2
\end{array}
\right]
=
\frac{1}{2((\xi,\xi)(\eta,\eta)-(\xi,\eta)^2)}
\left[
\begin{array}{ll}
(\eta,\eta)&-(\xi,\eta)\\
-(\xi,\eta)&(\xi,\xi)
\end{array}
\right]
\left[
\begin{array}{l}
(\xi,\xi)+1\\
(\eta,\eta)+1
\end{array}
\right].
\end{equation}

\subsection{Geodesics between preimage and image points on orbit surfaces.}

First we compute the geodesic distance of the preimage and image points
under class representatives eq. \ref{g1} from their scalar products.

{\bf 7 Lemma}: 
The length of all geodesic sections for 
fixed hyperbolic $g_0=g_0(c,\gamma)$ and orbit parameter $\rho$ is given by
\begin{eqnarray}
\label{g27}
&&\langle L(g_0(-c/2,-\gamma/2) x(\rho,\phi),
L(g_0(c/2,\gamma/2)x(\rho,\phi)\rangle
\\ \nonumber
&&= \cosh(2c)+(\cosh(2c)-\cos(2\gamma))(\sinh(2\rho))^2 \geq \cosh(2c).
\end{eqnarray}
The minimal geodesic length under
$L(g_0(c,\gamma))$ is reached for points on the straight geodesic orbit 
line with representative point $\rho=0,\; x=(1,0,0,0)$.

{\em Proof}: The scalar product between the  points $\lambda=\pm 1$ on the 
orbit line by evaluation of eq. \ref{g21} yields eq. \ref{g27}.
This expression depends on the parameters $c, \gamma$ of $g_0$ and on
the parameter $\rho$ which characterizes the orbit but is 
independent of the parameters $\lambda,\phi$ of the starting point
on the orbit.
The geodesic distance takes its minimum value 
$\cosh(2c)$ for $\rho=0$. Its length is determined by
the character $\chi(g_0)$ eq. \ref{g2}.

Next we characterize the geodesics between the preimage and image 
points. We particularize the general construction of 
lemma 6 to the geodesic between 
$\xi(\lambda),\; \lambda= -1,+1$. Both vectors have the same length.
From eqs. \ref{g23}, \ref{g25} it can be shown that this geodesic is a section 
on the circle $S^1(q(-1,+1),R(-1,+1))$ with center and radius
\begin{eqnarray}
\label{g28}
q(-1,+1)
&=& {\rm cotanh}(2\rho)\frac{\cosh(c)}{\cos(\gamma)}
(\cos(2\phi),\sin(2\phi),0),
\\ \nonumber
R(-1,+1)&=& \sqrt{(q(-1,+1)),q(-1,+1))-1}.
\end{eqnarray}

In Fig. 1 we show the orbit lines and geodesic sections for the 
Lorentz transformation $C_1= L(g_0(c,\gamma))$ which is 
one generator of the uniform lattice $\Gamma(M)$ studied in 
section 6. For the orbit parameters we use
six values ${\rm arccosh}(2\rho)=(0.0,0.2,0.4,0.6,0.8,1.0)$,
$\phi=0$.
In Fig. 2 we show the orbit and geodesic lines between
two vertices of the Weber-Seifert dodecahedron analyzed in 
sections 5,6.

\begin{center}
\includegraphics{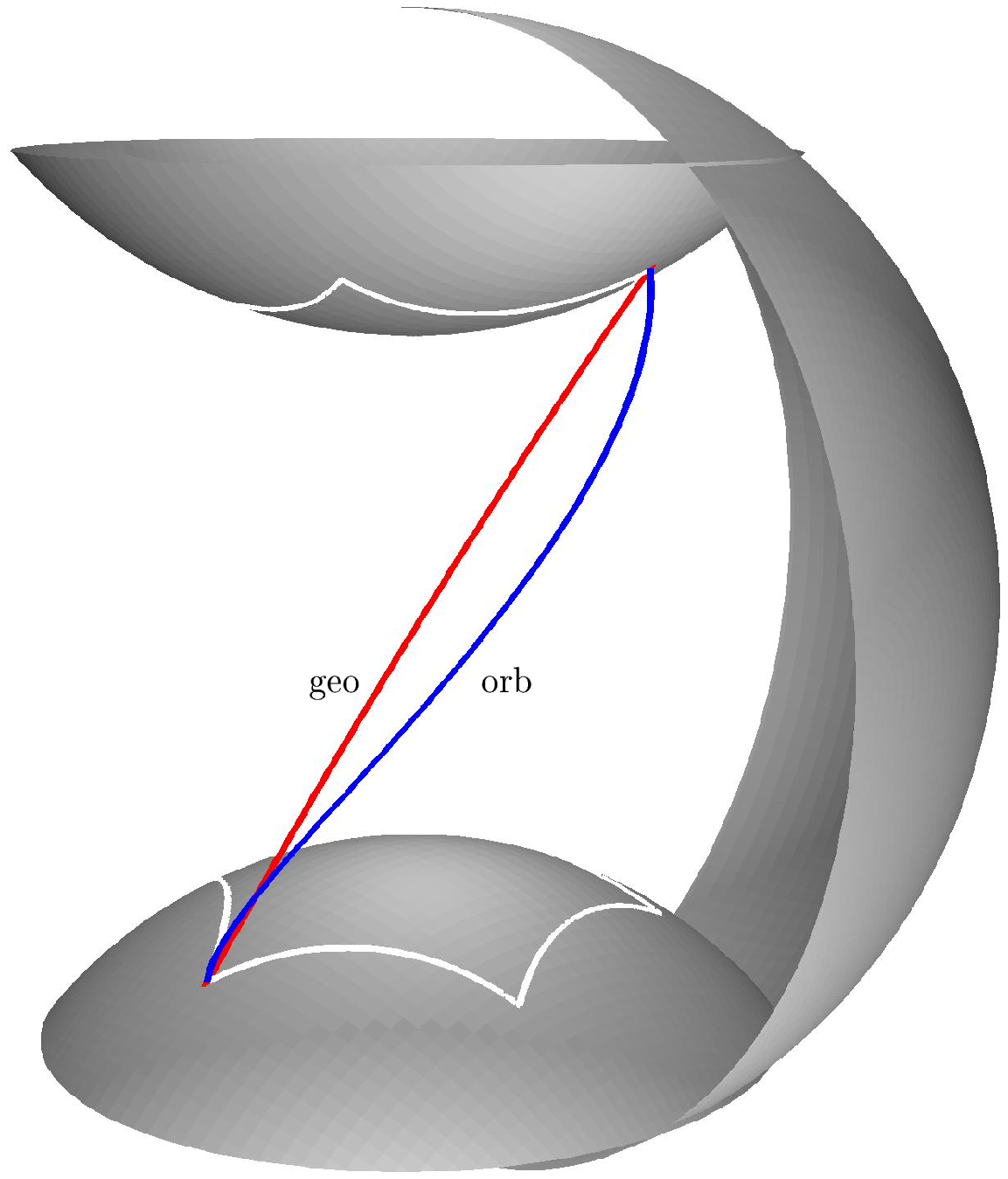}
\end{center}

Fig. 2. View of part of $B^3$ and the same two M\"obius spheres
as in Fig. 1. An orbit line (orb) connects  vertices  of opposite
pentagonal faces of the Weber-Seifert dodecahedron, the 
geodesic section between 
them is marked (geo).

\subsection{Defect angle of geodesic loops.}

Let a  geodesic loop start at a point $P \in M$ and intersect itself
at $P$. Denote by $\Delta$ the defect angle between the starting geodesic 
and its continuation after return and intersection. For a smooth  
geodesic loop we must have $\Delta=0$. 
We compute
$\Delta$ on the universal covering $B^3$ by considering along with
$P$ the preimage and image $g_0^{-1}P, g_0P$ under a fixed diagonal 
element  $g_0 \in \Gamma(M)$ eq. \ref{g1}
with parameters $c, \gamma$.  
As orbit representative $P$ we choose on the plane $\xi_3=0$ from 
eq. \ref{g23}
the point $\xi(\lambda) \in B^3, \lambda=0$ with $\phi=0$. 
The images of this plane under
$g_0^{-1}$ and $g_0$ are two M\"obius spheres similar to the ones used 
in eq. \ref{g24} but now for the parameter values 
$\lambda=\mp 2$. 
Consider the geodesic which runs on $B^3$ from
$\xi(0)$ to $\xi(2)=g_0\xi(0)$. When mapped to the compact manifold
$M$ with $g_0 \in \Gamma(M)$ , the geodesic intersects itself at $\xi(0)$.
To compare the directions at start and after intersection, we apply
$g_0^{-1}$. The image of the geodesic now runs
from $\xi(-2)$ to the intersection at $\xi(0)$ and so corresponds 
to $\lambda=(-2,0)$. The map $g_0$ is conformal, and so
we can measure the defect angle on $B^3$ between the directions of the
starting geodesic $\lambda=(0,2)$ and the continuation of the
geodesic $\lambda=(-2,0)$ at $\xi(0)$.

With the abbreviations
\begin{eqnarray}
\label{g29}
&&u:=\cosh(2c),\; v:=\sinh(2c),\; c:=\cos(2\gamma),\; s:=\sin(2\gamma), 
\\ \nonumber
&& \tau:=\cosh(2\rho),\; \sigma:=\sinh(2\rho),
\end{eqnarray}
we compute from 
eq. \ref{g23} the three points  
$\xi(-2), \xi(0), \xi(2)$  in terms of the parameters of
eq. \ref{g29},
\begin{equation}
\label{g30}
\xi(-2) = (1+\tau u)^{-1}
\left[
\begin{array}{l}
\sigma c\\
\sigma s\\
-\tau v
\end{array}
\right],\; 
\xi(0) = (1+\tau)^{-1}
\left[
\begin{array}{l}
\sigma \\
0\\
0
\end{array}
\right],\;  
\xi(2) = (1+\tau u)^{-1}
\left[
\begin{array}{l}
\sigma c\\
-\sigma s\\
\tau v
\end{array}
\right]. 
\end{equation}
All three points are on a single orbit of the continuous normalizer.
We observe from 
eq. \ref{g30} that, by the planar rotation 
\begin{eqnarray}
\label{g31}
&&R_{2,3}=
\left[
\begin{array}{ll}
c'&s'\\
-s'&c'
\end{array}
\right],
\\ \nonumber
&& c':=\tau v/\omega,\; s':=\sigma s/\omega,\; \omega:=\sqrt{\sigma^2s^2+\tau^2v^2}, 
\end{eqnarray}
applied to the components $2,3$ of all three vectors, their
new coordinate expressions become

\begin{equation}
\label{g32}
\xi(-2) = (1+\tau u)^{-1}
\left[
\begin{array}{l}
\sigma c\\
0\\
-\omega
\end{array}
\right],\; 
\xi(0) = (1+\tau)^{-1}
\left[
\begin{array}{l}
\sigma \\
0\\
0
\end{array}
\right],\;  
\xi(2) = (1+\tau u)^{-1}
\left[
\begin{array}{l}
\sigma c\\
0\\
\omega
\end{array}
\right], 
\end{equation}
and so all three vectors are in the single new $1,3$ plane of $B^3$.

We construct from 
eq. \ref{g25} the two geodesics which run  between the pairs of points
corresponding to $\lambda=(0,2)$ and $\lambda=(-2,0)$
respectively. Clearly both
geodesic sections run on the new $1,3$ plane and intersect at $\xi(0)$. 

Evaluating 
eqs. \ref{g25}, \ref{g26} for pairs of points we find for the center vectors 
$q(0,2),\;q(-2,0)$ of the
two geodesic circles 
\begin{eqnarray}
\label{g33}
&&q(0,2)=\mu_2 \xi(0)+\mu_1 \xi(2),
\\ \nonumber
&&q(-2,0)=\mu_1 \xi(-2)+\mu_2 \xi(0),
\\ \nonumber
&& \mu_1=(1+\tau u)\frac{\tau(u-c)}{\omega^2},\;
\mu_2=(1+\tau)\frac{\tau[-\sigma^2(cu-1)+\tau^2v^2]}
{\sigma^2 \omega^2}.
\end{eqnarray}

The radius vectors $\xi(0)-q(0,2),\xi(0)-q(-2,0)$
are perpendicular respectively to the starting and returning geodesics at 
$\xi(0)$. Therefore their angle of intersection determines the defect angle 
$\Delta$. From eqs. \ref{g32}, \ref{g33} we find in the new coordinates
\begin{equation}
\label{g34}
\xi(0)-q(-2,0)=
\left[
\begin{array}{l}
-\sigma^{-1}\\
0\\
\tau(u-c)/\omega
\end{array}
\right],\; 
\xi(0)-q(0,2)=
\left[
\begin{array}{l}
-\sigma^{-1}\\
0\\
-\tau(u-c)/\omega
\end{array}
\right].
\end{equation}
The two vectors differ only by a reflection in the new 3-coordinate axis
and so we get for half the defect angle the expression
\begin{eqnarray}
\label{g35}
\tan(\frac{1}{2}\Delta)
&=& \frac{(\xi(0)-q(0,2))_3}{(\xi(0)-q(0,2))_1}
\\ \nonumber
&=& \sigma \frac{\tau(u-c)}{\omega}
\\ \nonumber
&=& \sinh(2\rho)\frac{\cosh(2\rho)(\cosh(2c)-\cos(2\gamma))}
{\sqrt{\sinh(2\rho)^2\sin(2\gamma)^2+\cosh(2\rho)^2\sinh(2c)^2}}.
\end{eqnarray}
In the final expression  we replaced the abbreviations
from eq. \ref{g29}. We analyze the terms in the last expression.
The exponential parameters for a fixed non-trivial hyperbolic  $g_0$ must
obey $(\cosh(2c)-\cos(2\gamma))>0,\; |\sinh(2c)|>0$. Therefore the factor
of $\sinh(2\rho)$ in eq. \ref{g35} is always different from zero. 
The value $\Delta=0$ 
of the defect angle for a smooth geodesic loop  enforces 
the unique orbit line with representative $\sinh(2\rho)=0 \rightarrow \rho=0$. 

{\bf 8 Lemma}: 
The defect angle $\Delta$ for geodesic loops on $M$ associated with
$g_0$ is given as the function of group and orbit parameters by 
eq. \ref{g35}. The only smooth geodesic loops
associated with $g_0$ are sections of hyperbolic length $2c$ 
fixed by $\chi(g_0)$. They run  on the infinite 
geodesic line $\xi=(0,0,\xi_3), -1\leq \xi_3 \leq 1,$ 
mapped from $B^3$ to $M$.

\subsection{Orbit surfaces and geodesics for general Lorentz transformations.}

So far we dealt only with the action of a class representative of 
a Lorentz transformation on general points of $H^3$. 
The general Lorentz transformation is given 
from eq. \ref{g4} by conjugation with $g_2g_1$
as $L(g)=L((g_2g_1)g_0(g_2g_1)^{-1})$. Let 
$c, \gamma$ again denote the exponential parameters of
$g_0$.

By conjugation with $g_2g_1$ we introduce the two-parameter
commutative general normalizer 
$N_g:
=(g_2g_1)N_{g_0} (g_2g_1)^{-1}$ 
which is conjugate to $SO(1,1,R)\times SO(2,R)$ 
and commutes with $g$. 
We let this normalizer  
act on $H^3$ or $B^3$ and obtain orbit surfaces. 
Lemma 4 is  easily generalized to these orbit surfaces.
The geometric results given in Lemma 5-8 for $L(g_0)$ 
can be transcribed  to $L(g)$ 
if before we pass on
$H^3, B^3$ from initial coordinates $x$ to new ones
defined as $y=L((g_2g_1)^{-1})x$. 
For general $L(g)$ it  follows from lemma 7 
that the shortest geodesic loop
under $L(g)$ are sections of length fixed by the character $\chi(g)$  
on the image of a straight infinite geodesic  line
under the inverse map $L(g_2g_1)$ acting on the coordinates $y$.

This action is isometric and conformal.
Therefore the  expressions eqs. \ref{g27} for the 
geodesic length and eq. \ref{g35} for the defect angle remain true in terms
of the parameters for the diagonal form $g_0$ of $g$.
For given $g$ there is a unique set of shortest and smooth geodesics.

\section{The hyperbolic Coxeter group.}

The Weber-Seifert manifold is related to a hyperbolic
Coxeter group. It generates the dodecahedral tesselation and
has the uniform lattice $\gamma(M)$ as  a subgroup. The hyperbolic Coxeter 
group produces discrete symmetries of the Weber-Seifert manifold $M$.
Symmetries of compact manifolds play an important part in their
classification, see \cite{LE} pp. 266-279.

\subsection{The Coxeter group on $M(1,3)$ and $B^3$.}
There is a hyperbolic Coxeter group which has the uniform lattice $\Gamma(M)$
of the Weber-Seifert space as a subgroup. This Coxeter group 
\cite{RA} p. 284 has
the Coxeter  diagram\\ 
$\;\circ \frac{5}{}\circ\frac{3}{}\circ\frac{5}{}\circ\;$. Its
four generators $R_1,\ldots, R_4$ have the relations
\begin{eqnarray}
\label{g36}
&&R_1^2=R_2^2=R_3^2=R_4^2=e,
\\ \nonumber
&&(R_1R_2)^5=(R_2R_3)^3=(R_3R_4)^5=e,
\\ \nonumber
&&R_1R_3=R_3R_1,\;R_1R_4=R_4R_1,\; R_2R_4=R_4R_2.
\end{eqnarray}
The defining representation of this Coxeter group has a 
fundamental simplex in $M(1,3)$. Its boundaries are perpendicular
to four Weyl unit vectors $a_1,\ldots, a_4$ which generate the four
reflections $R_1,\ldots, R_4$. The dihedral angle between pairs of
Weyl reflection hyperplanes is related to the exponents 
$m_{12}=5, m_{23}=3, m_{34}=5 $ in the 
second line of eq. \ref{g36} by
\begin{equation}
\label{g37}
\langle a_i, a_{i+1} \rangle = \cos(\frac{\pi}{m_{i,i+1}}).
\end{equation}
A set of space-like unit Weyl vectors in $M(1,3)$ for
$\;\circ \frac{5}{}\circ\frac{3}{}\circ\frac{5}{}\circ\;$
is found as:
\begin{eqnarray}
\label{g38}
a_1 &=& (0,0,1,0),
\\ \nonumber
a_2 &=& (0,\frac{1}{2}\sqrt{-\tau+3},-\frac{1}{2}\tau,0),
\\ \nonumber
a_3 &=&  (0,-\sqrt{\frac{\tau+2}{5}}, 0, -\sqrt{\frac{-\tau+3}{5}}),
\\ \nonumber
a_4 &=& (\frac{1}{2}\sqrt{4\tau-1}, 0, 0,\frac{1}{2}\sqrt{4\tau+3}).
\end{eqnarray}
The Coxeter group acts on the hyperbolic space $H^3$ and on the conformal
ball $B^3$. In the conformal ball model we place 
the intersection of the first three Weyl hyperplanes planes at $\xi=0$.
The first three Weyl reflection hyperplanes in $B^3$ become planes (spheres 
of infinite radius) perpendicular to the space part of the
Weyl vectors in eq. \ref{g38}. The fourth Weyl hyperplane becomes a 
M\"obius sphere 
$S^2(q, R)$ which from eqs.  \ref{g15}, \ref{g38} has
\begin{equation}
\label{g39}
q=(0,0,\sqrt{\frac{4\tau+3}{4\tau-1}}),\; R=\frac{2}{\sqrt{4\tau-1}}.
\end{equation}  
The icosahedral Coxeter group 
${\bf\circ \frac{5}{}\circ\frac{3}{}\circ}$
is generated by $R_1,R_2,R_3$.
In $B^3$ this subgroup generates from the fundamental simplex 
the dodecahedron of Weber and Seifert \cite{WE}. From the Coxeter group action
it is formed by $120$ copies
of the fundamental Coxeter simplex. 
The unit vectors along the first three edge lines and axes of the fundamental 
simplex in $B^3$ are
\begin{eqnarray}
\label{g40}
e({\bf 5}) &=& (0,0,1),
\\ \nonumber
e({\bf 2})&=&(-\sqrt{\frac{-\tau+3}{5}},0,\sqrt{\frac{\tau+2}{5}}),
\\ \nonumber
e({\bf 3})&=&(-\sqrt{\frac{\tau+2}{3 \cdot 5}},
 -\frac{-\tau+3}{\sqrt{3\cdot 5}}, 
\sqrt{\frac{4\tau+3}{3 \cdot 5}}).
\end{eqnarray}
The $12$ outer faces of the 
dodecahedron are spherical pentagons on $12$ spheres $S^2(q,R)$
whose  center vectors $q$ are directed along the $12$ 5fold axes
of the dodecahedron, see Fig. 2. Inclusion of the generator $R_4$ generates 
a simplex tesselation of $B^3$. Sets of $120$ simplices form
the tiles of the dodecahedral tesselation of $B^3$.

\subsection{Preimage of the Coxeter group associated with $Sl(2,C)$}

In subsection 5.1  we gave on Minkowski space $M(1,3)$ the Weyl vectors which generate the 
hyperbolic Coxeter group and in subsection 6.2 the uniform lattice $\Gamma(M)$
 of the Weber-Seifert
hyperbolic dodecahedral manifold. For functions on $M(1,3)$ and on the coset
space $H^3$ it seems natural to include two-component spinor states
$\chi_1, \chi_2$ whose components transform according to $g \in Sl(2,C)$
while the functional dependence follows $L(g)$,
\begin{equation}
\label{g41}
T_g 
\left[
\begin{array}{l}
\chi_1(x)\\
\chi_2(x)
\end{array}
\right]
= g\;
\left[
\begin{array}{l}
\chi_1(L(g^{-1}x))\\
\chi_2(L(g^{-1}x))
\end{array}
\right].
\end{equation}
The operators eq. \ref{g41} form group homomorphisms,
\begin{equation}
\label{g42}
T_{g_1}T_{g_2}=T_{g_1g_2}.
\end{equation}
To handle such spinor states
we wish to extend $Sl(2,C)$
by preimages of Weyl
reflections. 
The action of $W_{e_2}$ on $M(1,3)$ can be expressed 
with respect to the matrix eq. \ref{g6} by
$\tilde{x}  \rightarrow \overline{\tilde{x}}$. Under $Sl(2,C)$
we find $g: \overline{\tilde{x}} 
\rightarrow \overline{g\ \tilde{x}g^{\dagger}}$.
Therefore we associate to the Weyl reflection  $W_{e_2}$ 
as preimage the automorphism
$C: g \rightarrow C g= \overline{g}$. In line with  eq. \ref{g10}
we extend this operator by conjugation to $g \circ C \circ g^{-1}$.
In operator products we use $C\circ g=\overline{g} \circ C$. 
First we find the preimage
of the Weyl operator $W_{e_3}$ as
\begin{eqnarray}
\label{g43}
g(e_3) \circ C \circ g(e_3)^{-1} &=& (g(e_3) \overline{g(e_3)}^{-1})\circ C
\\ \nonumber
&=& 
\left[
\begin{array}{ll}
0&i\\
i&0
\end{array}
\right] \circ C,
\\ \nonumber
g(e_3)&=&
\sqrt{\frac{1}{2}}\left[
\begin{array}{ll}
1& i\\
i&1
\end{array}
\right].
\end{eqnarray}
Next for any Weyl vector $a$ we can find a Lorentz transformation 
$L(l(a)),\;l(a) \in Sl(2,C):\; a= L(l(a))e_3 $. Then by 
extension of eq. \ref{g43} we get as preimage of the 
Weyl reflection $W_a$ by use of eqs. \ref{g9}, \ref{g10} the operator
\begin{eqnarray}
\label{g44}
s&=& l(a) g(e_3) \circ C \circ (l(a)g(e_3))^{-1}
\\ \nonumber 
&=& l(a) \left[
\begin{array}{ll}
0& i\\
i&0
\end{array}
\right]
\overline{l(a)}^{-1}
\circ C.
\end{eqnarray}

We now apply these results and eq. \ref{g44}
to find preimages $s_1,s_2,s_3,s_4$
associated with $Sl(2,C)$ for the generators of the Coxeter group.
We find
\begin{eqnarray}
\label{g45}
s_1&=& 
 -e \circ C,
\\ \nonumber
s_2&=&
\left[
\begin{array}{ll}
\exp(-i\pi/5)&0\\
0&\exp(i\pi/5)
\end{array}
\right] \circ C,
\\ \nonumber
s_3&=&
i\left[
\begin{array}{ll}
\sqrt{\frac{\tau+2}{5}}& -\sqrt{\frac{-\tau+3}{5}}\\
-\sqrt{\frac{-\tau+3}{5}}&-\sqrt{\frac{\tau+2}{5}}
\end{array}
\right] \circ C,
\\ \nonumber
s_4&=&
i\left[
\begin{array}{ll}
0&\frac{1}{2}(\sqrt{4\tau+3}+\sqrt{4\tau-1})\\
\frac{1}{2}(\sqrt{4\tau+3}-\sqrt{4\tau-1})&0
\end{array}
\right] \circ C.
\end{eqnarray}
From these expressions we compute the 
products $s_i\circ s_j$ and find
\begin{eqnarray}
\label{g46}
s_1 \circ s_1&=& s_2 \circ s_2=s_3 \circ s_3=s_4 \circ s_4=e,\\
\nonumber
s_1\circ s_2 
&=&
\left[
\begin{array}{ll}
-\exp(i\pi/5)&0\\
0&-\exp(-i\pi/5)
\end{array}
\right],\; 
(s_1\circ s_2)^5=e,
\\ \nonumber
s_2\circ s_3 
&=&
i\left[
\begin{array}{ll}
-\sqrt{\frac{\tau+2}{5}}\exp(-i\pi/5)& \sqrt{\frac{-\tau+3}{5}}\exp(-i\pi/5)\\
\sqrt{\frac{-\tau+3}{5}}\exp(i\pi/5)&\sqrt{\frac{\tau+2}{5}}\exp(i\pi/5)
\end{array}
\right], 
\\ \nonumber 
\frac{1}{2}{\rm Tr}(s_2\circ s_3)&=&-\frac{1}{2}=\cos(2\pi/3),\;
(s_2\circ s_3)^3=e,
\\ \nonumber
s_3\circ s_4 
&=&
\left[
\begin{array}{ll}
-\sqrt{\frac{-\tau+3}{5}}\frac{1}{2}(\sqrt{4\tau+3}-\sqrt{4\tau-1})
& \sqrt{\frac{\tau+2}{5}}\frac{1}{2}(\sqrt{4\tau+3}+\sqrt{4\tau-1})\\
-\sqrt{\frac{\tau+2}{5}}\frac{1}{2}(\sqrt{4\tau+3}-\sqrt{4\tau-1})&
-\sqrt{\frac{-\tau+3}{5}}\frac{1}{2}(\sqrt{4\tau+3}+\sqrt{4\tau-1})
\end{array}
\right],\; 
\\ \nonumber 
\frac{1}{2}{\rm Tr}(s_3\circ s_4)&=&-\frac{1}{2}\tau=\cos(4\pi/5),\;
(s_3\circ s_4)^5=e,
\\ \nonumber
s_1 \circ s_3&=-&s_3 \circ s_1,\; 
s_1 \circ s_4=-s_4 \circ s_1,\;
s_2 \circ s_4=-s_4 \circ s_2.
\end{eqnarray}
We use the trace relation  eq. \ref{g2} to determine the class parameters
of some products.
All the products eq. \ref{g46} of two generators are elements of $Sl(2,C)$.
By use of  $L(g)=L(-g)$ one finds that all relations eq. \ref{g46}
between  the preimages $s_i,\; i=1,\ldots, 4$ map correctly into the relations
eq. \ref{g36} of the Coxeter group.
We call the group generated by $\langle s_1,s_2,s_3,s_4\rangle$ 
the preimage of the Coxeter group 
$\;\circ \frac{5}{}\circ\frac{3}{}\circ\frac{5}{}\circ\;$.

\section{Homotopy, uniform lattice $\Gamma(M)$,   and homology groups of the hyperbolic dodecahedral space of 
Weber and Seifert.}

\subsection{The universal covering and its tesselation.}

Weber and Seifert \cite{WE} describe their closed hyperbolic dodecahedral
space $M$ as follows: Any face of the dodecahedron is glued
to its opposite face after a rotation by the angle $3\pi/5$.
The universal covering of $M$ must be a dodecahedral tesselation of
$H^3$. The tesselation condition at any vertex of the fundamental
dodecahedron enforces on $H^3$ the  dodecahedral tesselation found
from the Coxeter group. The uniform lattice $\Gamma(M)$ acts  
on the universal  covering $B^3$ by isometries and generates the 
dodecahedral tesselation. 
$\Gamma(M)$ must act without fixpoints and therefore
must be a proper subgroup of the Coxeter group
$\;\circ \frac{5}{}\circ\frac{3}{}\circ\frac{5}{}\circ\;$. 

\subsection{The uniform lattice $\Gamma(M)$.}

Following the description of Weber and Seifert we construct 
a first generator $C_1$ of $\Gamma(M)$. We enumerate six dodecahedral faces
by $i=1,\ldots, 6$ and
their  opposites by $\overline{i}$. We choose face $1$ perpendicular to
$e_3$ in $B^3$. The neighbour faces of face $1$ we enumerate counterclockwise as
$2,3,4,5,6$. The relation  of
\cite{WE} Fig.1 to the present notation is 
\begin{equation}
\label{g47}
\begin{array}{lllllll}
\cite{WE}:&A&B&C&D&E&F\\
{\rm present}:  &1&5&6&2&3&4
\end{array}
\end{equation}
The action of the icosahedral 
Coxeter group from eqs. \ref{g36}, \ref{g47} can now be given by 
signed permutations 
in cycle form.
The operation that moves face $\overline{1}$
to face  $1$ is a Lorentz transformation in the $(0,3)$ plane of $M(1,3)$
which commutes with a rotation by $2\tilde{\gamma}=3\pi/5$ 
in the $(1,2)$ plane, see eq. \ref{g8}. 
The parameter 
$2\tilde{c}$ of the
Lorentz transformation is found from the position of the Weyl hyperplane
perpendicular to $a_4$ eq. \ref{g38}. 
With this input and with eq. \ref{g8}
the generator $C_1$ is given  as
\begin{equation}
\label{g48}
C_1= L(g_0(\tilde{c},\tilde{\gamma})),\; \tilde{c}={\rm arccosh}(\frac{1}{2}\sqrt{4\tau+3}),
\;\tilde{\gamma}=3\pi/10.
\end{equation}
All other 
generators are conjugates of $C_1$ under the 
icosahedral Coxeter group and may be denoted as $C_i,\; i=1, \ldots, 6$.
Generators for opposite faces are inverses and so
\begin{equation}
\label{g49}
(C_i)^{-1}=C_{\overline{i}}.
\end{equation}
We are left with $6$ generators $C_i$ of $\Gamma(M)$. We shall 
employ the Coxeter group to find the relations between these 
generators. The first three Coxeter generators in the signed
cycle notation read
\begin{equation}
\label{g50}
R_1=(23)(46),\; R_2=(24)(56),\; R_3= (15)(2\overline{3}).
\end{equation}
Moreover we introduce for the 5fold rotation in the direction $i$ the
symbol $5_i, \; (5_i)^5=e$. In the cycle notation given above we have
$5_1=(23456)$. We also introduce the parity 
$P:=(1\overline{1})(2\overline{2})(3\overline{3})(4\overline{4})
(5\overline{5})(6\overline{6})$ which commutes with all elements of the
icosahedral Coxeter group. Then we find the following relation 
between the Coxeter generator $R_4$ and the generator $C_1$
of $\Gamma(M)$:
\begin{equation}
\label{g51}
R_4= C_1\;P\;5_1.
\end{equation}
As we have expressed generators of the Coxeter group by generators
of $\Gamma(M)$, we may rewrite relations of the 
Coxeter group in terms of generators of $\Gamma(M)$. We find
\begin{eqnarray}
\label{g52}
(R_3R_4)&=&C_5\; 5_2^{-2},\; (R_3R_4)^2=C_5\; C_{\overline{6}}\;5_2^{-4},
\\ \nonumber 
(R_3R_4)^3&=&C_5\; C_{\overline{6}}C_{\overline{3}}\;5_2^{-6},\; 
(R_3R_4)^4=C_5\; C_{\overline{6}}C_{\overline{3}}C_4\;5_2^{-8},
\\ \nonumber
(R_3R_4)^5&=&C_5\; C_{\overline{6}}C_{\overline{3}}C_4C_{\overline{1}}=e.
\end{eqnarray}
In the last step we applied $(5_2)^{10}=e$ and a relation of the Coxeter 
group from eq. \ref{g36}. The geometric origin of the relation 
between generators of $\Gamma(M)$ 
can now be seen: The powers of $(R_3R_4)$ generate  in $B^3$ 
5fold rotations around a fixed edge of the fundamental simplex and of 
the fundamental dodecahedron. The relations eq. \ref{g52} transcribe 
these reflection-generated rotations into relations between the generators of the 
uniform lattice $\Gamma(M)$. It is now easy to find the other relations. 
According to \cite{WE} there are six sets of equivalent edge lines
in the fundamental dodecahedron. Each of these sets gives 
a relation between the generators. The six relations become
\begin{eqnarray}
\label{g53}
&&C_{5}\; C_{\overline{6}}C_{\overline{3}}C_{4}C_{\overline{1}}=e,\;
C_{6}C_{\overline{2}}C_{\overline{4}}C_{5}C_{\overline{1}}=e,\;
C_{2}C_{\overline{3}}C_{\overline{5}}C_{6}C_{\overline{1}}=e,\;
\\ \nonumber
&&C_{3}C_{\overline{4}}C_{\overline{6}}C_{2}C_{\overline{1}}=e,\;
C_{4}C_{\overline{5}}C_{\overline{2}}C_{3}C_{\overline{1}}=e,\;
C_{4}C_{6}C_{3}C_{5}C_{2}=e.
\end{eqnarray}
{\bf 9 Lemma}: The uniform lattice  $\Gamma(M)$ of the Weber Seifert
hyperbolic dodecahedral manifold $M$ is generated by
$\langle C_1,C_2,C_3,C_4,C_5,C_6 \rangle$ with the 
relations eq. \ref{g53}. These relations are associated with 
$6$ edges $a,b,c,d,e,f$ of the dodecahedron in \cite{WE} Fig.1.

\subsection{$\Gamma(M)$ and the icosahedral Coxeter group.}

{\bf 10 Lemma}: The uniform lattice $\Gamma(M)$ of the Seifert-Weber
dodecahedral hyperbolic space forms a semidirect product 
with  the icosahedral Coxeter group
$\circ \frac{5}{}\circ\frac{3}{}\circ$,
\begin{equation}
\label{g54}
(\Gamma(M))\times_s (\circ \frac{5}{}\circ\frac{3}{}\circ).
\end{equation}
This semidirect product is
isomorphic to the full hyperbolic Coxeter group.

{\em Proof}: (i): Elements of
the icosahedral Coxeter subgroup generated by $R_1,R_2,R_3$
clearly by conjugation map generators of $\Gamma(M)$ into generators.
The elements of $\Gamma(M)$ have no fixpoint whereas any 
element of $\circ \frac{5}{}\circ\frac{3}{}\circ$ has the
point $\xi=0$ as fixpoint.
Therefore the intersection obeys
$(\Gamma(M))\cap (\circ \frac{5}{}\circ\frac{3}{}\circ)=e$.
These two properties suffice to show that the two groups form a 
semidirect product group 
eq. \ref{g54} with $\Gamma(M)$ the normal subgroup.
(ii): The remaining generator $R_4$ of the full hyperbolic
Coxeter group may be expressed from eq. \ref{g51}
as an element of the semidirect product group. Since 
then all four
generators eq. \ref{g36} of the full Coxeter group are in the semidirect
product, it must be isomorphic to the full Coxeter group,
\begin{equation}
\label{g55}
(\Gamma(M)) \times_s (\circ \frac{5}{}\circ\frac{3}{}\circ)\;
\sim \;\circ \frac{5}{}\circ\frac{3}{}\circ\frac{5}{}\circ .
\end{equation}
The isomorphism eq. \ref{g54} yields a new interpretation 
of the group relations eq. \ref{g53} for $\Gamma(M)$. If the 
Coxeter group is rewritten as the semidirect product with the normal
subgroup $\Gamma(M)$, the Coxeter relations enforce on $\Gamma(M)$ these 
group relations. 

The semidirect product group eq. \ref{g54} is a natural hyperbolic 
counterpart of a Euclidean symmorphic space group, with $\Gamma(M)$
being a non-commutative version of the translation group and 
$\circ \frac{5}{}\circ\frac{3}{}\circ$ of the point group.
It follows that a preimage of $\Gamma(M)$ in $Sl(2,C)$ can be  constructed 
from the generators given in section 5.2. 
The generator $C_1$ was expressed already in eq. \ref{g51}
by an element $g_0(c,\gamma) \in Sl(2,C)$. All elements
of $\Gamma(M)$ are even words in the generators of the Coxeter
group with preimages in $Sl(2,C)$.

From eq. \ref{g54}, the uniform lattice  $\Gamma(M)$ of the Weber-Seifert
manifold is a normal subgroup of the hyperbolic Coxeter group.
A conjugation with an element of the Coxeter group maps elements 
of $\Gamma(M)$ 
into one another and therefore relates the corresponding sets 
of geodesic loops beyond $\Gamma(M)$. 

The elements of
$\Gamma(M)$ are words in the $6$ generators $C_i$. These words must be
analyzed in view of the $6$ relations eq. \ref{g53}. An alternative
is to rewrite the words of $\Gamma(M)$ as new words in the $4$
generators $R_j$. The relations eq. \ref{g36} between these 
generators are much simpler to control, see \cite{HU} p. 171,
and so yield an efficient approach to the word problem.
By the methods of section 5.2, this analysis can be carried out
on the level of $Sl(2,C)$ without use of Lorentz transformations.

\subsection{The homology group $H_1(M)$.}
The homology group of the Weber-Seifert dodecahedral manifold
$M$ can be derived by abelianization 
of  $\pi_1(M) \sim \Gamma(M)$. Applying abelianization to the 
generators and to their relations eq. \ref{g44} one obtains:

{\bf 11 Lemma}: The homology group $H_1(M)$ as abelianization of the homotopy 
group $\Gamma(M)$ is
the direct product  
\begin{equation}
\label{g56}
H_1(M) = (Z/5Z) \times (Z/5Z) \times (Z/5Z).
\end{equation}
of three cyclic groups of order $5$. The icosahedral Coxeter group 
acts on these cyclic group with a modular representation by $3 \times 3$
matrices of integer entries modulo 5.\\
{\em Proof}: We apply abelianization to the defining relations
eq. \ref{g53}. The abelian image of a generator we denote by a label $a$.
If we multiply the first five relations with one another
and abelianize we find
\begin{equation}
\label{g57}
(C_1^a)^5= e.
\end{equation} 
For each generator $C_i^a$ we can find $5$ relations (or their inverses)
proportional to $C_i^a$. Multiplying them yields the analog of equation 
eq. \ref{g57} for $i=2,\ldots ,6$. It is possible to select $3$ 
independent generators
say $C_1^a,C_4^a,C_5^a$  
of order $5$ and to express the other ones by them in the form
of monomials,
\begin{eqnarray}
\label{g58}
C_j^a& =&  (C_1^a)^{m_{1j}} (C_4^a)^{m_{4j}} (C_5^a)^{m_{5j}},\; j=2,3,6,
\\ \nonumber
\left[ m_1,m_4,m_5\right]
&=& \left[
\begin{array}{lll}
1&2&2\\
3&2&4\\
3&4&2
\end{array}
\right].
\end{eqnarray}
where all integers are taken modulo $5$.
We can now implement a conjugation  action of the icosahedral
Coxeter group.
We write the conjugation of the three generators of $H_1(M)$ 
with the help of eq. \ref{g58} as 
\begin{eqnarray}
\label{g59}
R_i (C_l^a)R_i^{-1}
&=& (C_1^a)^{m_{1l}^i} (C_4^a)^{m_{4l}^i} (C_5^a)^{m_{5l}^i},\\
\nonumber
m^1&=& 
\left[
\begin{array}{lll}
1&2&0\\
0&4&0\\
0&2&1
\end{array}
\right],\;
m^2= 
\left[
\begin{array}{lll}
1&1&2\\
0&3&4\\
0&3&2
\end{array}
\right],\;
m^3= 
\left[
\begin{array}{lll}
0&0&1\\
0&1&0\\
1&0&0
\end{array}
\right],\; l=1,4,5.
\end{eqnarray}
The entries of all matrices in eqs. \ref{g58}, \ref{g59} are taken 
modulo $5$.
The three modular matrices $m^1,m^2,m^3$ obey $(m^l)^2=1_3,\; l=1,2,3$.
They generate a modular representation of the icosahedral Coxeter group.
Modular representations of the symmetric group are discussed 
in \cite{RO}.

\subsection{Conclusion: geodesic loops on $M$.}

Our main results on geodesic loops are:\\  
(1) Any $g \in \Gamma(M) \subset Sl(2,C)$ 
produces a variety  of geodesic loops.
The character $\chi(g)$ determines the class parameters, the
matrix $(g_2g_1)$ of eq. \ref{g4} that transforms $g$ to 
diagonal form $g_0$ yields 
the in-class  parameters of $g$.\\
(2) The continuous commutative normalizer $N_g \subset Sl(2,C)$ of the 
discrete subgroup generated by $g$ determines
orbits on $B^3$. Geodesic loops 
start and end   on the same orbit. Their length
is given by eq. \ref{g27}, the  defect angle at their self-intersection 
by eq. \ref{g35} in terms of the class and orbit 
parameters.\\
(3) For any $g$ there is a unique set of
shortest and smooth geodesic loops. All of them  are sections on a single
infinite geodesic line on $B^3$.\\
(4) For elements in the same class of $\Gamma(M)$,
the in-class parameters of the diagonalizing matrix $(g_2g_1)$ 
determine the normalizer $N_g$ as
a subgroup conjugate to $N_{g_0}$,  the conjugate orbits,
and the single infinite geodesic line for the shortest and
smooth geodesic loops.\\
(5) To compare geodesics arising from different classes of $\Gamma(M)$
one must study the words in this group. 
Symmetries, exemplified by the Weber-Seifert dodecahedral 
manifold, yield additional relations between geodesic loops.
For the Weber-Seifert manifold, the words can be rewritten in terms
of generators for the hyperbolic Coxeter group. Since the latter group has
the simpler relations eq. \ref{g36}, the word problem is simplified
by this rewriting.

\subsection*{Acknowledgment.}
The author owes many thanks to Linus Kramer, U. W\"urzburg, Germany
for comments and suggestions.

\newpage

\end{document}